\documentclass[
	a4paper, 
	10pt, 
	unnumberedsections, 
	twoside, 
]{LTJournalArticle}

\runninghead{} 

\footertext{\textit{RISC-V Summit Europe, Bologna, 8-12th June 2026}} 

\title{Code size reduction by advanced \\ near addressing modes} 

\author{%
	Kajetan Nürnberger\textsuperscript{1}, Thomas Röcker\textsuperscript{1} Gergely Fütő\textsuperscript{2}, Gábor Spaits\textsuperscript{2}, Horst Lehser\textsuperscript{2}
}

\date{\footnotesize\textsuperscript{\textbf{1}}Infineon Technologies AG \\ \textsuperscript{\textbf{2}}2HighTec EDV-Systeme GmbH}

\renewcommand{\maketitlehookd}{%
	\begin{abstract}
		\noindent To enable debugging and calibration of real time systems, which are in interaction with the real plant, the software used on those systems often has a huge number of global variables. The huge number of global variables exceed the range addressable relative to the global pointer. Therefore, addressing these variables normally needs two instructions. Other CPU architectures commonly used in the real time control systems domain address these by various near addressing modes. This results in significant code size reductions and performance boost. This paper discusses different variants to add such near addressing features to the RISC-V ISA. The impact on the code size is evaluated with different representative workloads.
	\end{abstract}
}

\begin{document}

\maketitle 


\section{Introduction}

Code size was a huge concern in the early days of the RISC-V ISA. With the introduction of C and Zc extensions a lot of those concerns have been addressed \cite{Perotti2020HWSWAF}. Still there are ongoing activities like \cite{RVG122:2020}. To the best of our knowledge none of those covers a proposal similar to the one presented.

 In many benchmarks the code size achieved with the RISC-V ISA is on a similar level than for other ISAs. For applications in the real time control domain there is still a gap for certain applications. Software developed for real time control systems has partially different patterns than commonly used SW. The development cycle requires that the SW is also monitored in detail during the interaction with the real plant. This is done by monitoring various variables. In addition to the pure monitoring, configuration parameters are also changed to calibrate the system. But this is only possible for global variables in the SW. So, in contrast to what is normally good practice in SW development a lot of global variables are contained in these systems. 
 
To enhance the addressing of global variables RISC-V offers the relaxing with respect to the global pointer (GP). As the immediate values between upper (20-bit) and lower (12-bit) immediate are not symmetric, the range addressable by GP is with 4 KB rather limited. Especially in bigger real time systems, the number of global variables is bigger than this space, in a range of hundreds of KB. Consequently, all variables exceeding this limited space need to be addressed by two instructions, with a load upper immediate before the actual load or store instruction. To address this, other architectures built for real time control applications, like TriCore™ or RH850, offer advanced near addressing features. Applying these has a significant impact on the code size for those architectures.

These architectures normally also offer multiple registers to store a base pointer for the near addressing. The reason for this is that in a physically addressed microcontroller there exists usually a fragmented memory map. Constants are normally directly accessed in the non-volatile memory (NVM). As there is usually a gap in the address map between NVM and random-access memory (RAM) it would not be possible to address constant and volatile variables with a single reference like GP. 

In the following sections additional load and store instructions to the RISC-V ISA are proposed, to get an enhanced addressing of global variables. Their impact on the overall code size is evaluated using a compiler prototype using representative real-world workloads.


\section{ISA Enhancement}

As the proposal is targeting real time workloads, we focus on the RV32 ISA only. The goal is to get load and store instructions with an increased immediate value. Therefore, the following instructions’ encoding would be affected:

\begin{itemize}
	\item LOAD
	\item STORE
	\item LOAD-FP
	\item STORE-FP
\end{itemize}

The goal of the proposed instructions is to stay within the RISC-V Instruction format and division of encoding space. Therefore, the proposed instructions are close to the original base instruction types. The opcode and funct3 field are kept and only the bits reserved for the source register are repurposed. Here two different possible enhancement versions are possible. The first one would be to keep one bit of the source register to be able to define two base pointers and use the remaining four bits as additional immediate value bits. This would result in two relative addressable ranges. Both of those ranges would have a size of 64 KB each. The second option would repurpose all the source register bits as additional immediate bits with a fixed register as base reference. This results in a single address range of 128 KB size.
For the load instructions signed and unsigned variants need to be considered. So here five additional instructions are required. It is not possible to fit these additional five instructions into the funct3 field of the load opcode. Therefore, an additional opcode is required for those. For the store instructions no signed versions are required so here it would still be possible to fit those additional operations into the already existing opcodes within additional encodings in the funct3 field. Similar for the floating-point load and store operations.

\section{Evaluation Method}

The benefit of the proposed enhancement obviously depends on the use of global variables in SW. Therefore, real applications are used for evaluation. The code footprint of these applications is in the range of half a megabyte. So, it is larger than commonly used embedded benchmarks. But it is still smaller by a factor of 10 to 50 in comparison to real complex modern automotive embedded applications. Normally with the code footprint of an embedded application also the RAM consumption is increased. For classic benchmarks where the code size is in up to the hundred KB range, it is often possible to fit the whole RAM in a GP relatively addressable range. The proposed extensions will not bring a benefit here. 
The impact on the code size is demonstrated with a prototype extension added to a RISC-V LLVM compiler. Here it is possible to generate the proposed instructions based on a threshold value for individual variable size. Variables below this threshold will be addressed with respect to the reference pointer by the proposed instructions. This is a common approach transferred from traditional real time compute architectures. To have a finer grain allocation here often also certain attributes for variable definition are available which are not yet included in the prototype. But as GP based addressing in RISC-V is normally used by advanced link time relaxation, also these methods can be transferred for the near addressing. As reference pointers currently the registers t0 and t1 are reserved.

\section{Results and Discussion}

To evaluate the impact of the proposed extensions, the different applications are compiled with and without the proposed extension. This is done once only for volatile variables and once for variable and constant variables. This enables that the impact on the RAM and ROM variables can be evaluated independently.

\begin{table} 
       \caption{Relative Code Size achieved with near addressing }
       \centering
       \begin{tabular}{l l r}
               \toprule
               Benchmark & RAM only & RAM \& ROM  \\
               \midrule
               1 & 96.0 \% & 95.7 \% \\
               2 & 92.0 \% & 91.7 \% \\
               3 & 85.3 \% & 82.5 \% \\
               \bottomrule
       \end{tabular}
       \label{tab:distcounts}
\end{table}
First the results show that the benefit of the proposed extension really depends on the way of coding. The first benchmark has still a high number of globally defined data but a lot of those are organized in structures. Here the benefit is limited as structure members are normally addressed relative to the structure start. So, in these cases a redirection is applied anyhow. The other two applications use more individual variables and here the benefit of the proposed extension is clearly visible. For the last application if both RAM and ROM near addressing is applied a code size reduction of almost 20 \% is reached. This is a significant enhancement. The magnitude of improvement can be observed also on more traditional CPU architectures for real time systems. 
C / Zc established an excellent basis to bring RISC-V to similar code-density as other architectures. However, for the case of automotive MCUs the suggested extensions enable further significant improvements and consequently positively impact related cost-functions. The proposed near addressing reduces next to the code size also the number of executed instructions. So, applying the near addressing also might also have a benefit in the execution time. A detailed study of this is scope of future work.
Another observation which is visible on the above results is that the impact on the ROM data is significantly less than on the RAM data. This raises the question how valid the request for two independent data ranges which are near addressable is? Within the ROM memory, often significantly bigger data structures like look up tables are located. This gives an indication that the results above will generalize and a single data range might be giving similar benefits. Especially in case of big applications where also the RAM requirements are in the range of hundreds of kilobytes to megabytes. In case a single range is used also the impact on the overall RISC-V ecosystem is less. 
In case of a single range, the GP can be reused as a base pointer, and the extension would be without an impact on the ABI. In case of two independent ranges at least a second base register needs to be defined which would impact the ABI. For compilers a single data range can also be implemented by enhancing GP-based relaxation. 
Therefore, the recommendation is to go for a single register as reference and an enhanced addressable range of 128 KB. With the linker relaxation techniques, this also can overcome the cumbersome appliance of near addressing modes for SW developers and make RISC-V to the superior ISA with respect to code size in the embedded real time domain.



\printbibliography 


\end{document}